\documentclass[a4paper,fleqn,usenatbib]{mnras}

\usepackage[T1]{fontenc}
\usepackage{ae,aecompl}

\usepackage{graphicx}	
\usepackage{amsmath}	
\usepackage{amssymb}	

\title[ArH$^+$ in the Crab]{Modelling the ArH$^+$ emission from the Crab Nebula}

\author[F. D. Priestley et al.]{
F. D. Priestley,$^{1}$
M. J. Barlow$^{1}$
and S. Viti$^{1}$
\\
$^{1}$Department of Physics and Astronomy, University College London, Gower Street, London WC1E 6BT, UK\\
}

\date{Accepted XXX. Received YYY; in original form ZZZ}

\pubyear{2017}

\begin{document}
\label{firstpage}
\pagerange{\pageref{firstpage}--\pageref{lastpage}}
\maketitle

\begin{abstract}
We have performed combined photoionization and photodissociation region (PDR) modelling of a Crab Nebula filament subjected to the synchrotron radiation from the central pulsar wind nebula, and to a high flux of charged particles; a greatly enhanced cosmic ray ionization rate over the standard interstellar value, $\zeta_0$, is required to account for the lack of detected [C I] emission in published Herschel SPIRE FTS observations of the Crab Nebula. The observed line surface brightness ratios of the OH$^+$ and ArH$^+$ transitions seen in the SPIRE FTS frequency range can only be explained with both a high cosmic ray ionization rate and a reduced ArH$^+$ dissociative recombination rate compared to that used by previous authors, although consistent with experimental upper limits. We find that the ArH$^+$/OH$^+$ line strengths and the observed H$_2$ vibration-rotation emission can be reproduced by model filaments with $n_{\rm{H}} = 2 \times 10^4$ cm$^{-3}$, $\zeta = 10^7 \zeta_0$ and visual extinctions within the range found for dusty globules in the Crab Nebula, although far-infrared emission from [O I] and [C II] is higher than the observational constraints. Models with $n_{\rm{H}} = 1900$ cm$^{-3}$ underpredict the H$_2$ surface brightness, but agree with the ArH$^+$ and OH$^+$ surface brightnesses and predict [O I] and [C II] line ratios consistent with observations. These models predict HeH$^+$ rotational emission above detection thresholds, but consideration of the formation timescale suggests that the abundance of this molecule in the Crab Nebula should be lower than the equilibrium values obtained in our analysis.
\end{abstract}




\section{Introduction}
\label{sec:intro}

Argonium (ArH$^+$), the first noble gas molecule discovered in space, was detected by \citet{barlow2013} in emission in the J=1-0 and J=2-1 transitions in the Crab Nebula, along with an OH$^+$ transition at 971 GHz. The Crab Nebula is a young supernova remnant (SNR), consisting of a central pulsar wind nebula (PWN) and a surrounding network of filaments ionised by the PWN synchrotron emission \citep{hester2008}. Molecular hydrogen (H$_2$) \citep{graham1990,loh2010,loh2011,loh2012} emission had previously been detected in the Crab Nebula, with modelling by \citet{richardson2013} suggesting it originates from a trace molecular component in mostly atomic gas, shielded from ionising radiation at the centre of the filaments. \citet{barlow2013} noted that ArH$^+$ is principally formed by the reaction Ar$^+$ + H$_2$ $\to$ ArH$^+$ + H \citep{roach1970}, while the main destruction mechanism is ArH$^+$ + H$_2$ $\to$ Ar + H$_3^+$ \citep{schilke2014}.

Since its discovery in the Crab Nebula, ArH$^+$ has been found to be a fairly common molecule in the interstellar medium (ISM). \citet{schilke2014} reported ArH$^+$ J=1-0 absorption at a range of velocities along several galactic sightlines, while \citet{muller2015} detected ArH$^+$ absorption from a foreground galaxy in the spectrum of a lensed blazar. \citet{schilke2014} found the interstellar ArH$^+$ abundance to be uncorrelated with the abundances of the molecular gas tracers HF and H$_2$O$^+$, while chemical modelling of ArH$^+$ in the ISM by \citet{schilke2014} and \citet{neufeld2016} showed that it traces gas where the hydrogen is almost entirely atomic. ArH$^+$ was predicted to peak in abundance at the edges of interstellar clouds, where the ultraviolet (UV) radiation field is strongest.

The Crab Nebula is a substantially different environment to the interstellar clouds considered by \citet{schilke2014}. The PWN is a strong source of synchrotron radiation, including X-rays, which behave very differently to the UV radiation in the interstellar case. Additionally, the filaments must be permeated by the high energy charged particles which produce the synchrotron radiation, equivalent to but much more intense than the cosmic rays included in ISM astrochemical modelling. Both can provide an additional source of Ar$^+$, potentially allowing more ArH$^+$ to be formed, while the strong radiation field and generally harsh environment would tend to destroy molecules. This paper will attempt to model the chemistry of ArH$^+$ under the conditions present in the Crab Nebula with the goal of reproducing the observed emission.

\section{Description of \textsc{ucl\_pdr}}
\label{sec:code}

In this work we make use of the substantially upgraded 1D \textsc{ucl\_pdr} photodissociation region (PDR) code. It was first developed by \citet{bell2005,bell2006} and further augmented by \citet{bayet2011}. The code assumes a semi-infinite slab geometry, and the chemistry and thermal balance are calculated self-consistently at each depth point into the slab, producing chemical abundances, emission line strengths and gas temperatures as a function of depth. The gas temperature is calculated by taking into account a wide range of heating and cooling processes (see \citet{bell2006} and \citet{rollig2007}, where the code has also been benchmarked against other codes). The gas can be heated by the photoejection of electrons from dust grains and polycyclic aromatic hydrocarbons (PAHs), FUV pumping, photodissociation of H$_2$ molecules, cosmic rays, exothermic reactions, C I photoionization, H$_2$ formation, gas-grain collisions and turbulence. The gas is mainly cooled through emission from collisionally excited atoms and molecules and by interactions with the cooler dust grains. The original code included cooling from [O I], [C I] and [C II] fine structure lines and CO rotational lines (up to J=11--10). \citet{bayet2011} improved the \textsc{ucl\_pdr} code by adding the cooling, and hence the calculation of the line brightnesses, from $^{13}$CO, C$^{18}$O, CS and HCN transitions. The line brightnesses for all the coolant species are calculated using the Large Velocity Gradient (LVG) approximation (see e.g. \citet{vandertak2007}). All the collisional rates are taken from the Leiden Atomic and Molecular Database \citep{schoier2010}. The chemical network is user-specified, using the rate-coefficient formalism of the UMIST database \citep{mcelroy2013}.

The \textsc{ucl\_pdr} code has now been further improved by the following additions, made by T. Bell:
\begin{enumerate}
\item LVG calculation of line intensities (and cooling contributions) from arbitrary new species. This requires appropriate collisional data to be available.
\item Two-sided FUV illumination, either in the form of scaled Draine fields or full radiation SEDs which can be take from a radiative transfer model.
\item Photoreaction rates are now calculated from available cross-section data when full input radiation SEDs are provided.
\item A detailed treatment of the H$_2$ formation rate based on laboratory data, either in the form of a rate equation \citep{cazaux2004} or combined with results from advanced Monte Carlo rate calculations \citep{iqbal2012}.
\item The possibility to run gas-grain models including freeze-out, surface reactions, and various desorption mechanisms.
\item Implementation of the X-ray chemistry and physics, following \citet{meijerink2005} and \citet{stauber2005}. The shape and intensity of the incident X-ray spectrum can be specified, as appropriate for either YSOs or AGN.
\item Inclusion of a limited PAH chemistry, following \citet{wolfire2003,wolfire2008}.
\item Inclusion of an updated vibrationally excited H$_2$ chemistry and vibrational de-excitation heating rate calculation \citep{rollig2006}.
\item Fully parallel OpenMP implementation, with good scaling efficiency (the speed-up scales well with increasing number of cores).
\end{enumerate}
The user is now able to specify most model parameters within input text files, including dust extinction properties, input SEDs, cloud density structure, collisional data for LVG calculations, photorate cross sections, and so on. \textsc{ucl\_pdr} will shortly be made publically available from https://uclchem.github.io.

\section{Method}
\label{sec:method}

We have modelled a single Crab Nebula filament using a combination of two codes: \textsc{mocassin}, a Monte Carlo photoionisation code \citep{ercolano2003,ercolano2005,ercolano2008}, and the upgraded version of \textsc{ucl\_pdr} described above. \citet{owen2015} used \textsc{mocassin} to model the global properties of the Crab Nebula, fitting the predicted SED and line strengths to multi-wavelength observations in order to determine the dust and gas properties for a given distribution of matter. We used the dust and gas densities, dust grain size distribution and the elemental composition of the gas from their clumpy model VI, assuming amorphous carbon dust grains. We adopted a shell of matter with these properties located at 2.5 pc from a central point source for modelling with \textsc{mocassin}, using the Crab SED from \citet{hester2008} and the total SED luminosity of $1.3 \times 10^{38}$ erg s$^{-1}$ adopted by \citet{owen2015}. The shell thickness was adjusted so that the transition between ionised and neutral gas occurred at the outer edge, which required a thickness of $3.5 \times 10^{16}$ cm. The output SED was integrated between 912 \r{A} and 2400 \r{A} to give the transmitted UV field longward of the Lyman limit - $31$ Draines (see \citet{draine1978}) - and from 0.1 \r{A} to 100 \r{A} to give an X-ray flux of $0.35$ erg cm$^{-2}$ s$^{-1}$.

\begin{figure}
  \centering
  \includegraphics[width=\columnwidth]{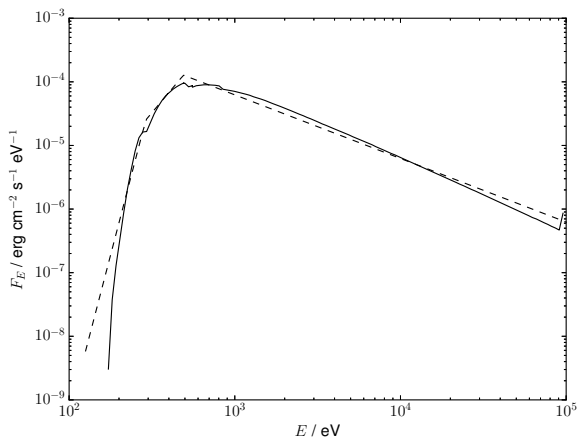}
  \caption{The output X-ray spectrum from \textsc{mocassin} (solid line; see text) with the three component power law fit used by \textsc{ucl\_pdr} (dashed line).}
  \label{fig:xrayspec}
\end{figure}

The filament was modelled in \textsc{ucl\_pdr} as a one dimensional slab, with the UV and X-ray fluxes that had been determined from \textsc{mocassin} incident on one side of the slab. The hydrogen nucleus number density, $n_{\rm{H}}$, was initially set to 1900 cm$^{-3}$, the value found for the clumps in model VI of \citet{owen2015}, while the initial cosmic ray ionisation rate was left as the standard interstellar value ($\zeta_0 = 1.3 \times 10^{-17} \rm{s}^{-1}$). The gas phase elemental abundances are listed in Table~\ref{tab:abun}. To account for the conditions in the Crab Nebula, \textsc{ucl\_pdr} was modified to include the effect of the Crab's higher dust-to-gas mass ratio ($\frac{M_d}{M_g} = 0.031$; \citet{owen2015}) on the H$_2$ formation and heating rates. The cosmic ray heating rate was taken from the code \textsc{cloudy} \citep{ferland1998}, as the default rate in \textsc{ucl\_pdr} (from \citet{goldsmith2001}) is for dark clouds, and may not be appropriate for the Crab Nebula, while the X-ray heating rate was calculated using the heating efficiency for pure helium, the most abundant element in the Crab, taken from \citet{dalgarno1999}. The X-ray absorption cross-sections were changed to those appropriate for the Crab Nebula abundances, and the input X-ray spectrum was approximated by a three component power law fit to the \textsc{mocassin} output spectrum, as shown in Figure~\ref{fig:xrayspec}.

\begin{table}
  \centering
  \caption{Gas phase elemental abundances, relative to hydrogen, used in PDR modelling, taken from model IV of \citet{owen2015}.}
  \begin{tabular}{cccc}
    \hline
    Element & Abundance & Element & Abundance \\
    \hline
    H & $1.00$ & N & $2.5 \times 10^{-4}$ \\
    He & $1.85$ & O & $6.2 \times 10^{-3}$ \\
    C & $1.02 \times 10^{-2}$ & Ar & $1.0 \times 10^{-5}$ \\
    \hline
  \end{tabular}
  \label{tab:abun}
\end{table}

ArH$^+$ was incorporated into an existing XDR chemical network using the reaction network from \citet{schilke2014}, with their photodissociation rate replaced by the one calculated by \citet{roueff2014} for the Crab Nebula, with the correction given by \citet{neufeld2016} (Ar + H$_2^+ \to$ ArH$^+$ + H, instead of Ar + H$_2^+ \to$ Ar$^+$ + H$_2$). Additional reactions to account for the effects of X-rays were included. The argon reactions incorporated and their rates are listed in Table~\ref{tab:xrayreac}. The rates of reactions 13 to 16 were calculated according to the methods detailed by \citet{meijerink2005}. The rates for reactions 17 and 18 were taken from \citet{shull1982} and from \citet{kingdon1996}, respectively. We also included reactions involving the H$^-$ anion from the UMIST database \citep{mcelroy2013}, to account for its role in the formation of H$_2$ in the Crab Nebula \citep{richardson2013}. Electron impact excitation rates for the first six rotational levels of ArH$^+$ were taken from \citet{hamilton2016}, and we used energy levels and Einstein A-coefficients from the Cologne Database for Molecular Spectroscopy (CDMS) \citep{muller2001,muller2005}, in order to calculate the emissivity of ArH$^+$ rotational transitions.

\begin{table*}
  \centering
  \caption{Argon reactions used for modelling the Crab nebula ArH$^+$ emission. CR refers to cosmic rays, CRPHOT to secondary photons produced by cosmic rays, XR to X-rays, XRSEC to secondary electrons produced by X-rays, XRPHOT to secondary photons from X-rays, h$\nu$ to a photon, $\zeta$ to cosmic ray or X-ray ionisation rates, $\sigma(E)$ is the X-ray ionisation cross section at energy $E$, $F(E)$ is the X-ray flux at energy $E$ and $\omega$ is the dust albedo.}
  \begin{tabular}{cccc}
    \hline
    & Reaction & Rate coefficient & Source \\
    \hline
    1 & Ar + CR $\to$ Ar$^{+}$ + e$^-$ & $10 \zeta_{\rm{H,cr}}$ & \citet{schilke2014} \\
    2 & Ar + CRPHOT $\to$ Ar$^+$ + e$^-$ & $0.8 \frac{\zeta_{\rm{H}_2\rm{,cr}}}{1 - \omega}$ & \citet{schilke2014} \\
    3 & Ar + H$_2^+$ $\to$ ArH$^+$ + H & $10^{-9}$ cm$^3$ s$^{-1}$ & \citet{schilke2014} \\
    4 & Ar + H$_3^+$ $\to$ ArH$^+$ + H$_2$ & $8 \times 10^{-10}$ exp$\left(\frac{6400 \rm{K}}{T}\right)$ cm$^3$ s$^{-1}$ & \citet{schilke2014} \\
    5 & Ar$^+$ + e$^-$ $\to$ Ar + h$\nu$ & $3.7 \times 10^{-12} \left(\frac{T}{300 \rm{K}}\right)^{-0.651}$ cm$^3$ s$^{-1}$ & \citet{schilke2014} \\
    6 & Ar$^+$ + H$_2$ $\to$ ArH$^+$ + H & $8.4 \times 10^{-10} \left(\frac{T}{300 \rm{K}}\right)^{0.16}$ cm$^3$ s$^{-1}$ & \citet{schilke2014} \\
    7 & ArH$^+$ + H$_2$ $\to$ Ar + H$_3^+$ & $8 \times 10^{-10}$ cm$^3$ s$^{-1}$ & \citet{schilke2014} \\
    8 & ArH$^+$ + CO $\to$ Ar + HCO$^+$ & $1.25 \times 10^{-9}$ cm$^3$ s$^{-1}$ & \citet{schilke2014} \\
    9 & ArH$^+$ + O $\to$ Ar + OH$^+$ & $8 \times 10^{-10}$ cm$^3$ s$^{-1}$ & \citet{schilke2014} \\
    10 & ArH$^+$ + C $\to$ Ar + CH$^+$ & $8 \times 10^{-10}$ cm$^3$ s$^{-1}$ & \citet{schilke2014} \\
    11 & ArH$^+$ + e$^-$ $\to$ Ar + H & $10^{-9}$ cm$^3$ s$^{-1}$ & \citet{schilke2014} \\
    12 & ArH$^+$ + h$\nu$ $\to$ Ar$^+$ + H & $4.20 \times 10^{-12}$ exp$\left(-3.27A_V\right)$ s$^{-1}$ & \citet{roueff2014} \\
    13 & Ar + XR $\to$ Ar$^{++}$ + e$^-$ + e$^-$ & $\int_{E_{\rm{min}}}^{E_{\rm{max}}} \sigma(E) \frac{F(E)}{E} \rm{d}E$ & \citet{verner1995} \\
    14 & Ar$^+$ + XR $\to$ Ar$^{++}$ + e$^-$ & $\int_{E_{\rm{min}}}^{E_{\rm{max}}} \sigma(E) \frac{F(E)}{E} \rm{d}E$ & \citet{verner1995} \\
    15 & Ar + XRSEC $\to$ Ar$^{+}$ + e$^-$ & $5.97 \zeta_{\rm{H,X-ray}}$ & \citet{lennon1988} \\
    16 & Ar + XRPHOT $\to$ Ar$^{+}$ + e$^-$ & $0.8 \frac{\zeta_{\rm{H}_2\rm{,X-ray}}}{1 - \omega}$ & \citet{schilke2014} \\
    17 & Ar$^{++}$ + e$^-$ $\to$ Ar$^+$ + h$\nu$ & $2.71 \times 10^{-11} \left(\frac{T}{300 \rm{K}}\right)^{-0.75}$ cm$^3$ s$^{-1}$ & \citet{shull1982} \\
    18 & Ar$^{++}$ + H $\to$ Ar$^+$ + H$^+$ & $10^{-15}$ cm$^3$ s$^{-1}$ & \citet{kingdon1996} \\
    \hline
  \end{tabular}
  \label{tab:xrayreac}
\end{table*}

\section{Results}
\label{sec:results}

\begin{figure}
  \centering
  \includegraphics[width=\columnwidth]{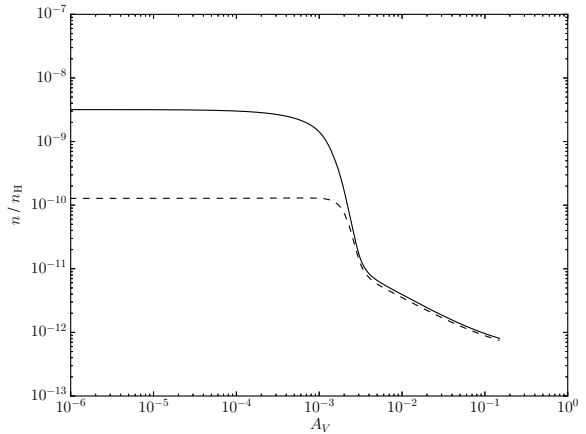}
  \caption{Ar$^+$ (solid line) and ArH$^+$ (dashed line) abundances relative to hydrogen nuclei as a function of $A_V$, for an interstellar cloud model using the same parameters as \citet{schilke2014}.}
  \label{fig:schilke}
\end{figure}

As a check, \textsc{ucl\_pdr} was run using the same parameters as for the model interstellar cloud studied by \citet{schilke2014}. Figure~\ref{fig:schilke} shows the resulting abundances of Ar$^+$ and ArH$^+$ as a function of the visual extinction depth, $A_V$. The abundances of both species are found to be slightly smaller than those in \citet{schilke2014}, but are within a factor of a few throughout the cloud, and show the same behaviour with increasing $A_V$. As shown by \citet{schilke2014}, the main factor affecting the Ar$^+$ and ArH$^+$ abundances is the molecular hydrogen fraction, which reacts with and destroys both species.

\begin{figure}
  \centering
  \includegraphics[width=\columnwidth]{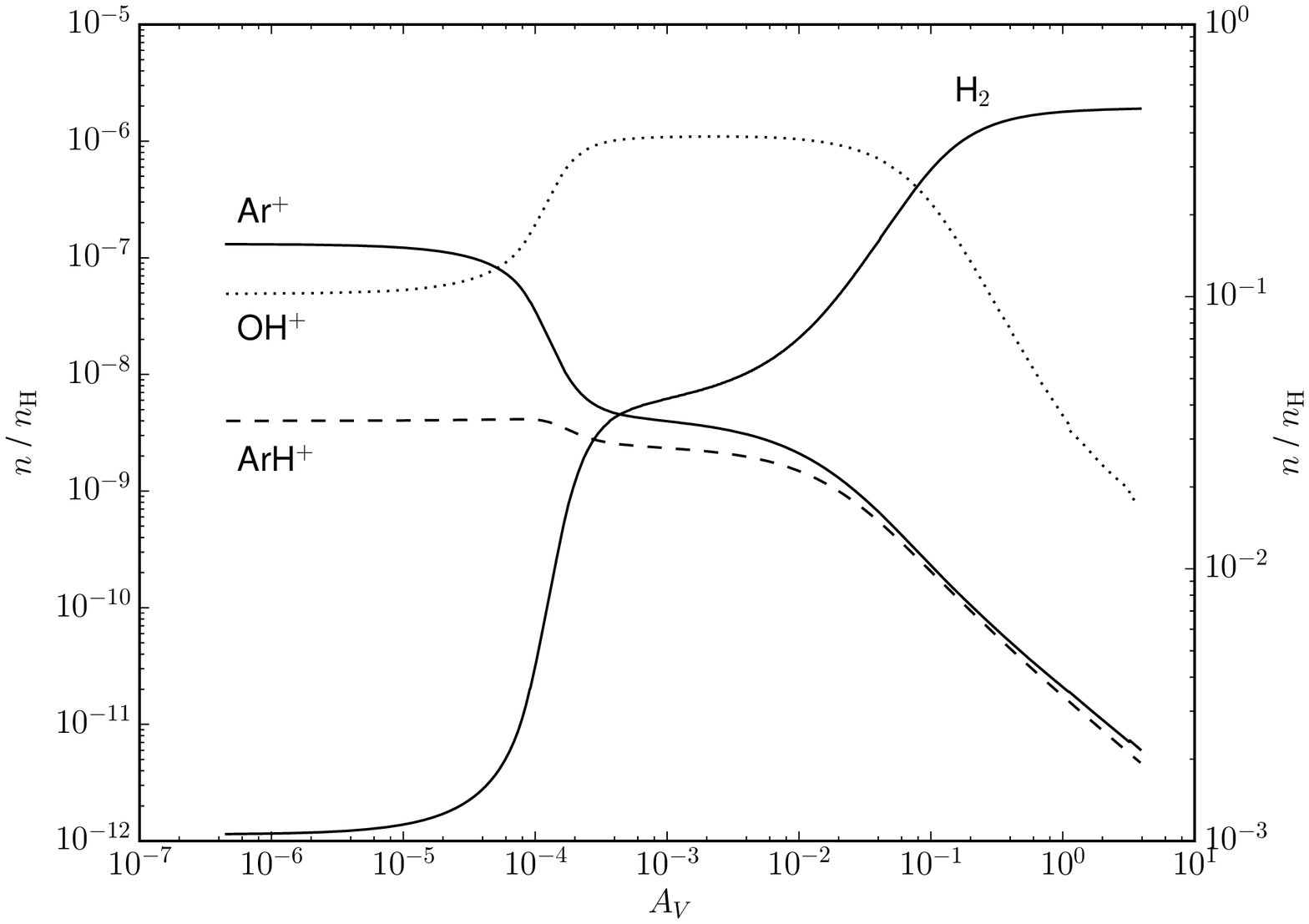}
  \caption{H$_2$, Ar$^+$, ArH$^+$ and OH$^+$ abundances as a function of $A_V$, for an initial Crab Nebula model. The H$_2$ abundance scale is on the right, the scale for the other species on the left.}
  \label{fig:crababun}
\end{figure}

Figure~\ref{fig:crababun} shows the abundances as a function of $A_V$ of H$_2$, Ar$^+$, ArH$^+$ and OH$^+$ (also observed by \citet{barlow2013}), for an initial Crab Nebula model clump. The model parameters used are listed in Table~\ref{tab:models}. The H$_2$ abundance is similar throughout the clump to the \citet{schilke2014} IS cloud model, but the abundance of ArH$^+$ is enhanced by over an order of magnitude, due to the additional production of Ar$^+$ by X-rays. The ArH$^+$ abundance is larger than or comparable to the peak interstellar cloud value in Figure~\ref{fig:schilke} for all but the highest $A_V$ values. The molecular ion OH$^+$, which has a similar formation mechanism (O$^+$, from cosmic ray or X-ray ionisation of neutral O, reacting with H$_2$), is also present at fairly high abundances, although its peak abundance is located at a higher $A_V$, and therefore deeper into the filament, than ArH$^+$.

\begin{table*}
  \centering
  \caption{Model parameters used for our initial Crab Nebula filament, and for our final models successfully reproducing the Herschel SPIRE FTS observations. $k$(e$^-$) is the rate coefficient of the ArH$^+$ + e$^-$ dissociative recombination reaction.}
  \begin{tabular}{cccccccc}
    \hline
    Model & $n_{\rm{H}}$ / cm$^{-3}$ & $\zeta$ / $\zeta_0$ & $F_{UV}$ / Draines & $F_X$ / erg cm$^{-2}$ s$^{-1}$ &  $A_V$ / mag & $A_V$/$N_{\rm{H}}$ / mag cm$^2$ & $k$(e$^-$) / cm$^3$ s$^{-1}$ \\
    \hline
    Initial & $1900$ & $1$ & $31$ & $0.35$ & $3.0$ & $6.289 \times 10^{-22}$ & $10^{-9}$ \\
    D3Z7 & $1900$ & $10^7$ & $31$ & $0.35$ & $0.05$ & $6.289 \times 10^{-22}$ & $10^{-11}$ \\
    D3Z8 & $1900$ & $10^8$ & $31$ & $0.35$ & $0.05$ & $6.289 \times 10^{-22}$ & $10^{-11}$ \\
    D4Z7 & $2 \times 10^4$ & $10^7$ & $31$ & $0.35$ & $0.01$ & $6.289 \times 10^{-22}$ & $10^{-11}$ \\
    D4Z8 & $2 \times 10^4$ & $10^8$ & $31$ & $0.35$ & $0.01$ & $6.289 \times 10^{-22}$ & $10^{-11}$ \\
    AVD3Z7 & $1900$ & $10^7$ & $31$ & $0.35$ & $0.30$ & $2.264 \times 10^{-21}$ & $10^{-11}$ \\
    AVD3Z8 & $1900$ & $10^8$ & $31$ & $0.35$ & $0.30$ & $2.264 \times 10^{-21}$ & $10^{-11}$ \\
    AVD4Z7 & $2 \times 10^4$ & $10^7$ & $31$ & $0.35$ & $0.30$ & $2.264 \times 10^{-21}$ & $10^{-11}$ \\
    AVD4Z8 & $2 \times 10^4$ & $10^8$ & $31$ & $0.35$ & $0.30$ & $2.264 \times 10^{-21}$ & $10^{-11}$ \\
    \hline
  \end{tabular}
  \label{tab:models}
\end{table*}

\begin{figure}
  \centering
  \includegraphics[width=\columnwidth]{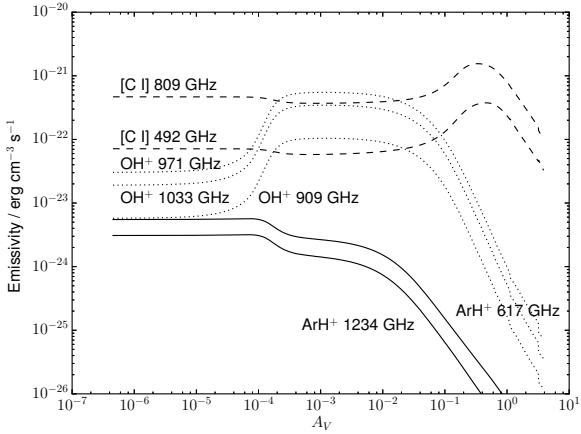}
  \caption{Emissivities of the strongest emission lines between 450 and 1550 GHz as a function of $A_V$ for the initial Crab Nebula model shown in Figure~\ref{fig:crababun}. ArH$^+$ transitions are shown as solid lines, OH$^+$ as dotted lines and neutral carbon as dashed lines.}
  \label{fig:crabemis}
\end{figure}

This initial Crab model predicts that significant amounts of ArH$^+$ and OH$^+$ should be formed in gas of the observed composition and density subjected to the attenuated radiation field from the Crab PWN, in agreement with the observations of \citet{barlow2013}. This model predicts a higher abundance of OH$^+$ than ArH$^+$. However, \citet{barlow2013} found that the ArH$^+$ rotational line emission was significantly stronger than that from OH$^+$, so we calculated the predicted emissivities for both species. Figure~\ref{fig:crabemis} shows the line emissivities as a function of $A_V$ for the strongest transitions predicted by this Crab Nebula model, in the 450 to 1550 GHz range covered by the Herschel SPIRE spectrometer. The predicted emissivities of two [C I] and three OH$^+$ lines are higher than those of the two ArH$^+$ lines in this frequency range throughout the cloud, and the ArH$^+$ J=1-0 617 GHz transition is predicted to have a higher emissivity than the J=2-1 line at 1234 GHz. In contrast, \citet{barlow2013} found the J=2-1 line to be much stronger than the J=1-0 line, and of the other predicted lines only the OH$^+$ 971 GHz transition was detected, at a comparable strength to the ArH$^+$ 617 GHz line.

To investigate this discrepancy, a grid of models was run having hydrogen number densities of $n_{\rm{H}} = 1900$, $2 \times 10^4$, $2 \times 10^5$ and $2 \times 10^6$ cm$^{-3}$, and cosmic ray ionisation rates of $\zeta = 1$, $10^4$, $10^5$, $10^6$, $10^7$ and $10^8$ $\zeta_0$. These enhanced values cover the limits estimated by \citet{richardson2013}, based on the observed Crab synchrotron radiation and the degree of magnetic trapping. The grid was run using the UV and X-ray fluxes from the attenuated PWN spectrum described above. To compare surface brightnesses consistently between models with different densities, the line emissivities were integrated up to $A_V = 3$ in each model.

\begin{figure}
  \centering
  \includegraphics[width=\columnwidth]{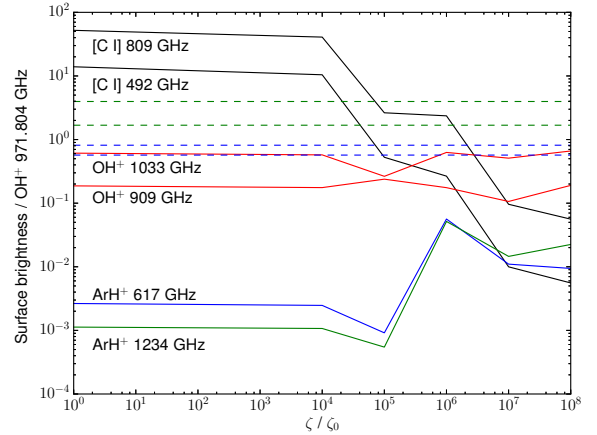}
  \caption{Line surface brightness ratios relative to OH$^+$ 971 GHz, of the other transitions of OH$^+$ (red), carbon (black) and ArH$^+$ J=1-0 (blue) and J=2-1 (green) from Figure~\ref{fig:crabemis} versus the cosmic ray ionisation rate $\zeta$, in units of the interstellar cosmic ray ionisation rate $\zeta_0 = 1.3 \times 10^{-17}$ s$^{-1}$, for models with $n_{\rm{H}} = 1900$ cm$^{-3}$. The dashed lines show upper and lower observational bounds for the ArH$^+$ transitions, from \citet{barlow2013}.}
  \label{fig:linescrabd2e3}
\end{figure}

\begin{figure}
  \centering
  \includegraphics[width=\columnwidth]{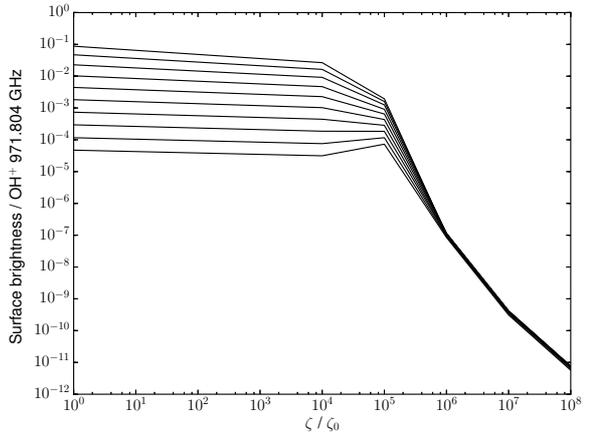}
  \caption{CO line surface brightness ratios relative to OH$^+$ 971 GHz of rotational transitions from J=4-3 (upper) to J=13-12 (lower) versus the cosmic ray ionisation rate $\zeta$, in units of the interstellar cosmic ray ionisation rate $\zeta_0 = 1.3 \times 10^{-17}$ s$^{-1}$, for models with $n_{\rm{H}} = 1900$ cm$^{-3}$.}
  \label{fig:colinesd2e3}
\end{figure}

Figure~\ref{fig:linescrabd2e3} shows the surface brightness of the various transitions plotted in Figure~\ref{fig:crabemis}, relative to the OH$^+$ 971 GHz transition, with each ratio plotted as a function of $\zeta$, for models with $n_{\rm{H}} = 1900$ cm$^{-3}$. The dashed lines show the observed range of ratios relative to OH$^+$ 971 GHz for the two ArH$^+$ transitions. The relative surface brightnesses of the ArH$^+$ lines are much lower than observed, with the highest values occuring for $\zeta = 10^6 \zeta_0$. The ratio between the two ArH$^+$ transitions is also in conflict with observation except at the highest values of $\zeta$. The non-detection of the neutral carbon lines also favours a high value of $\zeta$, as carbon atoms are converted to C$^+$ ions by cosmic ray ionisation, while the relative intensities of the OH$^+$ lines are generally compatible with observations. Figure~\ref{fig:colinesd2e3} shows line surface brightnesses relative to the OH$^+$ 971 GHz transition versus $\zeta$ for the ten CO rotational transitions in the SPIRE FTS frequency range (J=4-3 to J=13-12). All the CO transitions are predicted to be significantly weaker than the OH$^+$ lines, in agreement with the lack of detected CO emission. With increasing $\zeta$, the rate of destruction of CO by cosmic rays and the temperature both rise, resulting in lower surface brightnesses (due to a lower CO abundance) and converging line strengths (as higher J levels can more easily be populated through collisions).

\begin{figure}
  \centering
  \includegraphics[width=\columnwidth]{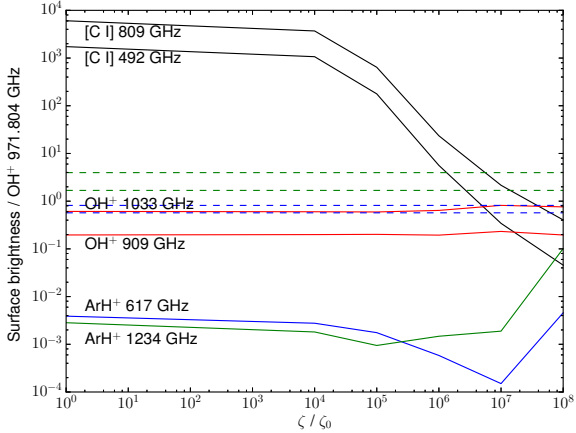}
  \caption{Line surface brightness ratios relative to OH$^+$ 971 GHz versus the cosmic ray ionisation rate $\zeta$ in units of $\zeta_0$, for models with $n_{\rm{H}} = 2 \times 10^6$ cm$^{-3}$. The dashed lines show upper and lower observational bounds for the ArH$^+$ transitions, from \citet{barlow2013}.}
  \label{fig:linescrabd2e6}
\end{figure}

Figure~\ref{fig:linescrabd2e6} shows the line surface brightness ratios versus $\zeta$ for the same transitions as plotted in Figure~\ref{fig:linescrabd2e3}, for a hydrogen nucleus density of $2 \times 10^6$ cm$^{-3}$. None of the enhanced density models reproduce the observed ArH$^+$/OH$^+$ surface brightness ratios adequately, while for $n_{\rm{H}} = 2 \times 10^6$ cm$^{-3}$ even the highest $\zeta$ models still predict significant [C I] emission. The strength of the two ArH$^+$ lines relative to each other is best reproduced by models with high ($\ge 10^7 \zeta_0$) cosmic ray ionisation rates, which also produce the surface brightness ratios closest to those observed for ArH$^+$, and weak enough [C I] emission to be consistent with observation. All models predict OH$^+$ 1033 GHz emission approximately equal in strength to the SPIRE-FTS ArH$^+$ 617 GHz line. While this OH$^+$ transition is not widely detected, in one detector (SSW B1) we did find an emission line at 1033 GHz with a slightly lower surface brightness than that of the OH$^+$ 971 GHz line in the same detector, which would be consistent with these results.

\begin{figure}
  \centering
  \includegraphics[width=\columnwidth]{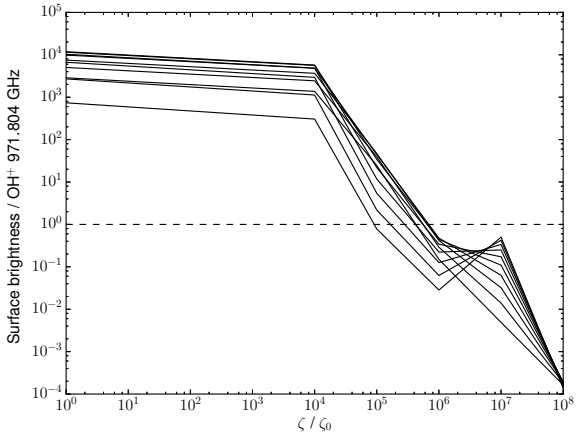}
  \caption{CO line surface brightness ratios relative to OH$^+$ 971 GHz of rotational transitions from J=4-3 (upper) to J=13-12 (lower) versus the cosmic ray ionisation rate $\zeta$ in units of $\zeta_0$, for models with $n_{\rm{H}} = 2 \times 10^6$ cm$^{-3}$.}
  \label{fig:colinesd2e6}
\end{figure}

Figure~\ref{fig:colinesd2e6} shows the relative surface brightnesses of the CO transitions versus $\zeta$ for a cloud density of $2 \times 10^6$ cm$^{-3}$. The CO emission becomes stronger at higher densities, although it remains much weaker than OH$^+$ 971 GHz at high $\zeta$. The lack of detected CO emission excludes models with lower values of $\zeta$ and densities above $2 \times 10^4$ cm$^{-3}$, with the maximum value considered ($10^8 \zeta_0$) required to produce low enough CO emission at all densities.

\begin{figure}
  \centering
  \includegraphics[width=\columnwidth]{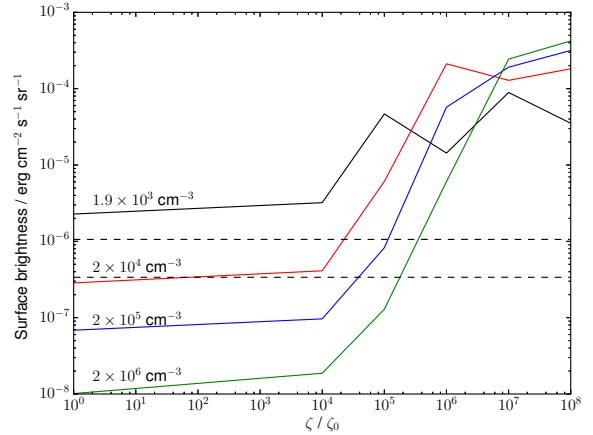}
  \caption{Absolute surface brightness integrated up to $A_V = 3$ of the OH$^+$ 971 GHz transition versus cosmic ray ionisation rate $\zeta$ for models with $n_{\rm{H}} = 1900$ cm$^{-3}$ (black), $2 \times 10^4$ cm$^{-3}$ (red), $2 \times 10^5$ cm$^{-3}$ (blue) and $2 \times 10^6$ cm$^{-3}$ (green). The dashed lines show upper and lower observational bounds from \citet{barlow2013}.}
  \label{fig:oh971av3}
\end{figure}

\begin{figure}
  \centering
  \includegraphics[width=\columnwidth]{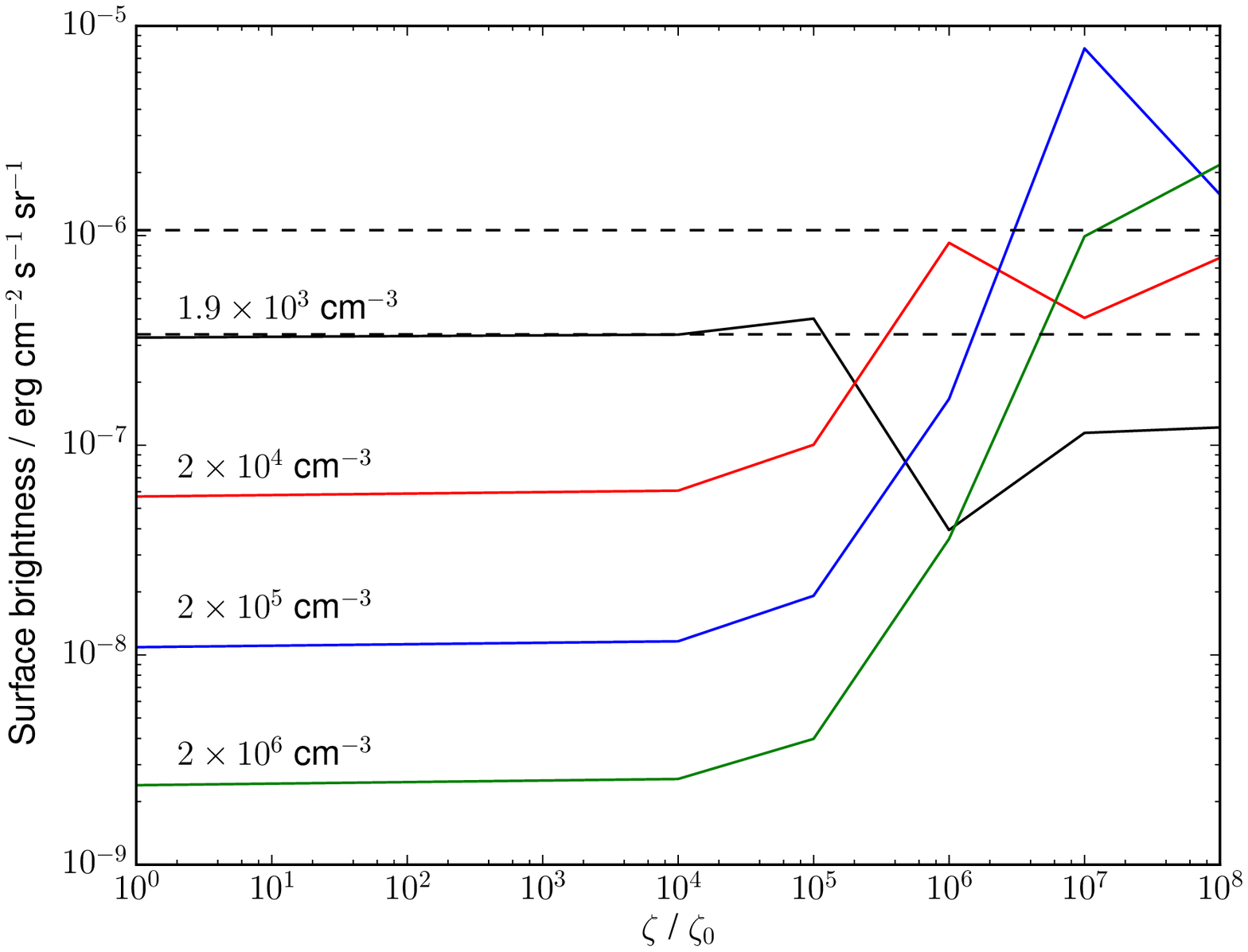}
  \caption{Absolute surface brightness integrated up to $A_V = 0.01$ of the OH$^+$ 971 GHz transition versus cosmic ray ionisation rate $\zeta$ for models with $n_{\rm{H}} = 1900$ cm$^{-3}$ (black), $2 \times 10^4$ cm$^{-3}$ (red), $2 \times 10^5$ cm$^{-3}$ (blue) and $2 \times 10^6$ cm$^{-3}$ (green). The dashed lines show upper and lower observational bounds from \citet{barlow2013}.}
  \label{fig:oh971av1e-2}
\end{figure}

\begin{figure}
  \centering
  \includegraphics[width=\columnwidth]{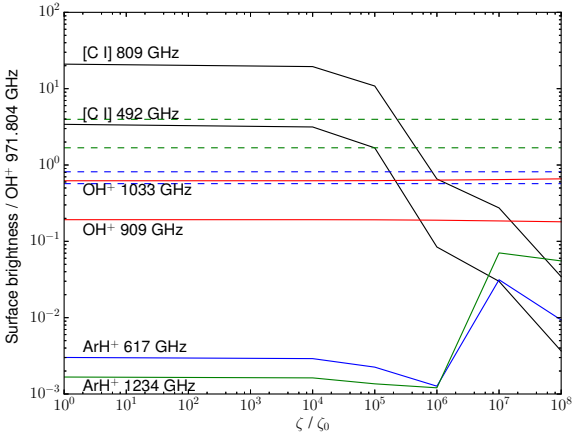}
  \caption{Line surface brightness ratios relative to OH$^+$ 971 GHz versus the cosmic ray ionisation rate $\zeta$ in units of $\zeta_0$, for models with $n_{\rm{H}} = 2 \times 10^4$ cm$^{-3}$, with the emissivity integrated up to $A_V = 0.01$, as for our D4Z7 and D4Z8 models (see Table~\ref{tab:models}). The dashed lines show upper and lower observational bounds for the ArH$^+$ transitions, from \citet{barlow2013}.}
  \label{fig:linescrabd2e4av1e-2}
\end{figure}

Figure~\ref{fig:oh971av3} shows the absolute surface brightness of the OH$^+$ 971 GHz transition as a function of $\zeta$, for the four densities investigated. \citet{barlow2013} found surface brightnesses of around $10^{-7}$ to $10^{-6}$ erg cm$^{-2}$ s$^{-1}$ sr$^{-1}$ for this line, much lower than the predicted values from our high $\zeta$ models, and for all our lowest density models. Figure~\ref{fig:oh971av1e-2} shows the surface brightness when the emissivity is integrated only up to a visual extinction of $A_V = 0.01$. In this case the $n_{\rm{H}} = 2 \times 10^4$ cm$^{-3}$ models have surface brightnesses closer to the observed values, with the three highest $\zeta$ models falling within the observed range. Figure~\ref{fig:linescrabd2e4av1e-2} shows line surface brightness ratios integrated up to $A_V = 0.01$ for a density of $2 \times 10^4$ cm$^{-3}$. Compared to the $A_V = 3$ case, the OH$^+$ and ArH$^+$ ratios are mostly unaffected, whereas there is a significant reduction in the predicted strength of the undetected [C I] transitions at low values of $\zeta$.

\begin{figure}
  \centering
  \includegraphics[width=\columnwidth]{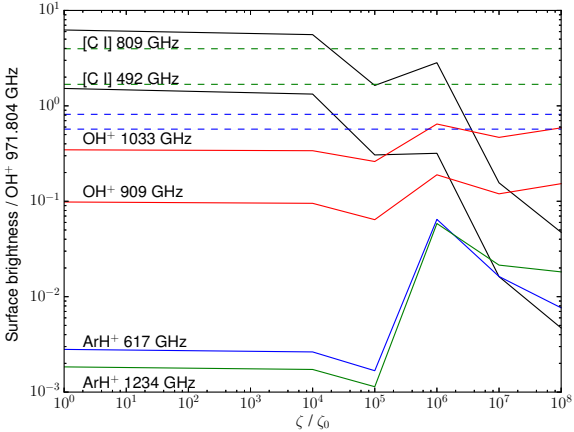}
  \caption{Line surface brightness ratios relative to OH$^+$ 971 GHz versus the cosmic ray ionisation rate $\zeta$ in units of $\zeta_0$, for models with $n_{\rm{H}} = 1900$ cm$^{-3}$, for the Crab Nebula PWN spectrum at a distance of 1 pc. The dashed lines show upper and lower observational bounds for the ArH$^+$ transitions, from \citet{barlow2013}.}
  \label{fig:lines1pcd2e3}
\end{figure}

A high value of $\zeta$ is required to account for the lack of neutral carbon and CO emission, but an increased X-ray/UV flux could also produce many of the same effects. To investigate this possibility, we repeated our \textsc{mocassin} modelling with the distance between the ionizing source and the filament reduced from $2.5$ pc to 1 pc. This resulted in input UV and X-ray field strengths of 94 Draines and 1.3 erg cm$^{-1}$ s$^{-1}$, respectively. Figure~\ref{fig:lines1pcd2e3} shows relative line surface brightnesses versus $\zeta$ for a cloud density of $1900$ cm$^{-3}$, using this increased radiation field. At low values of $\zeta$, the relative brightnesses of the neutral carbon lines are significantly reduced compared to the previous case, while the ArH$^+$ lines are increased slightly. However, to reduce the [C I] line surface brightnesses to levels consistent with observation, the cosmic ray ionization rate must still be at least $10^7 \zeta_0$, by which point the radiation field is no longer the dominant source of ionization and the two models give virtually identical results. We therefore conclude that an increased UV and X-ray radiation field cannot account for the observed emission line strengths, and high values of $\zeta$ are required.

The rate coefficient used for the ArH$^+$ + $e^-$ reaction, $10^{-9}$ cm$^3$ s$^{-1}$, is an upper limit found experimentally by \citet{mitchell2005} for low energy collisions. As this upper limit is comparable to the rate coefficient for the ArH$^+$ + H$_2$ reaction ($8 \times 10^{-10}$ cm$^3$ s$^{-1}$), if the electron density is similar to that of molecular hydrogen the destruction rate of ArH$^+$ may be overestimated. The grid was rerun with reduced electron recombination rates of $10^{-10}$ and $10^{-11}$ cm$^3$ s$^{-1}$ in order to investigate the effects on the abundance of ArH$^+$. The highest gas temperature reached in any of the PDR models is around $12000$ K, corresponding to a thermal energy of about $1.0$ eV, while \citet{mitchell2005} found the reaction rate for electron energies below $2.5$ eV was negligible.

\begin{figure}
  \centering
  \includegraphics[width=\columnwidth]{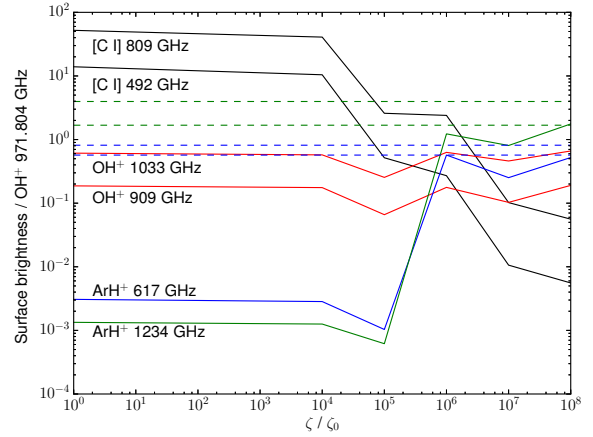}
  \caption{Line surface brightness ratios relative to OH$^+$ 971 GHz versus the cosmic ray ionisation rate $\zeta$ in units of $\zeta_0$, for models with $n_{\rm{H}} = 1900$ cm$^{-3}$ and a rate coefficient for ArH$^+$ + e$^-$ of $10^{-11}$ cm$^3$ s$^{-1}$. The dashed lines show upper and lower observational bounds for the ArH$^+$ transitions, from \citet{barlow2013}.}
  \label{fig:lineserate-11d2e3}
\end{figure}

\begin{figure}
  \centering
  \includegraphics[width=\columnwidth]{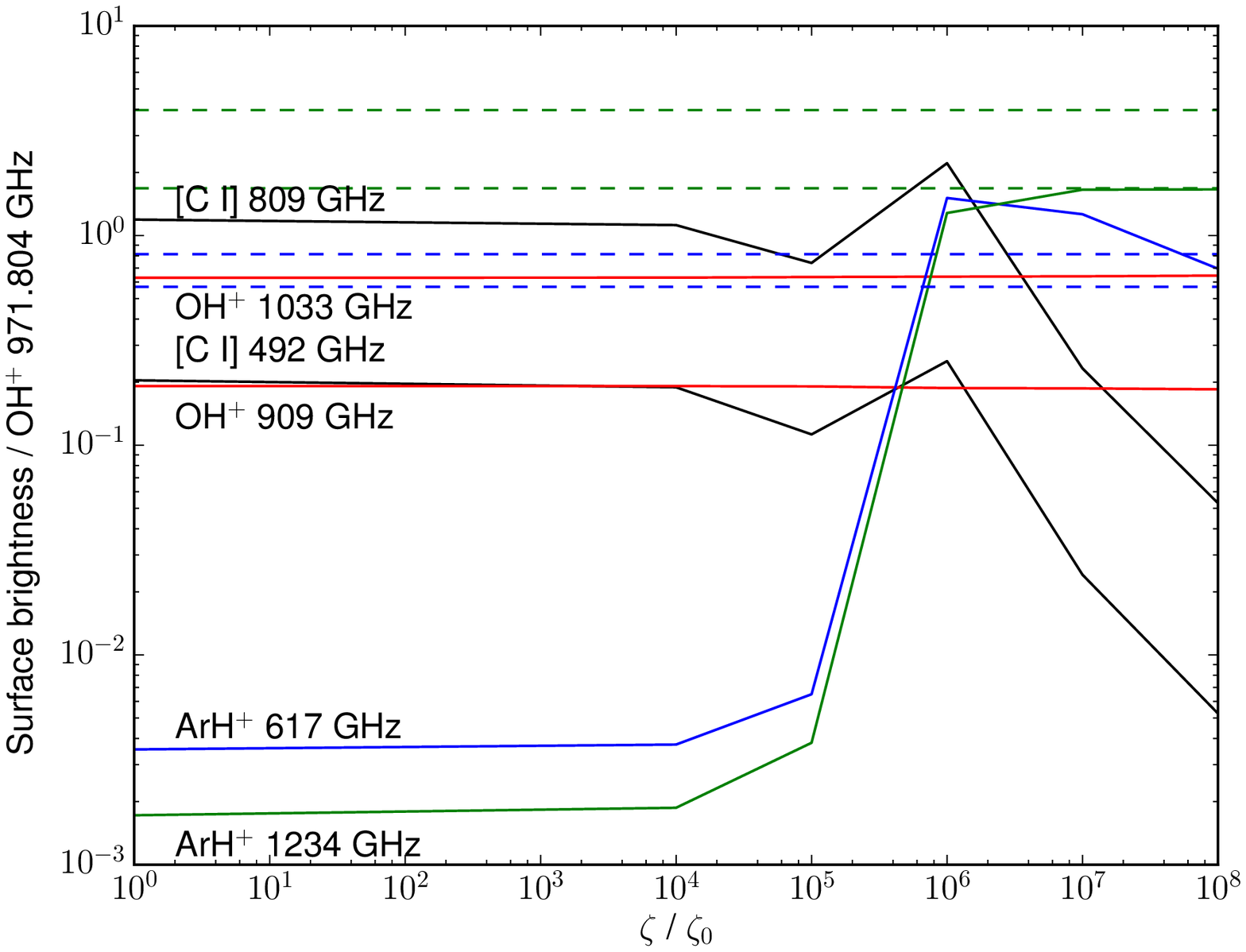}
  \caption{Line surface brightness ratios relative to OH$^+$ 971 GHz versus the cosmic ray ionisation rate $\zeta$ in units of $\zeta_0$, for models with $n_{\rm{H}} = 1900$ cm$^{-3}$ and a rate coefficient for ArH$^+$ + e$^-$ of $10^{-11}$ cm$^3$ s$^{-1}$, with the emissivity integrated up to $A_V = 0.05$, as for our D3Z7 and D3Z8 models (see Table~\ref{tab:models}). The dashed lines show upper and lower observational bounds for the ArH$^+$ transitions, from \citet{barlow2013}.}
  \label{fig:lineserate-11d2e3av5e-2}
\end{figure}

\begin{figure}
  \centering
  \includegraphics[width=\columnwidth]{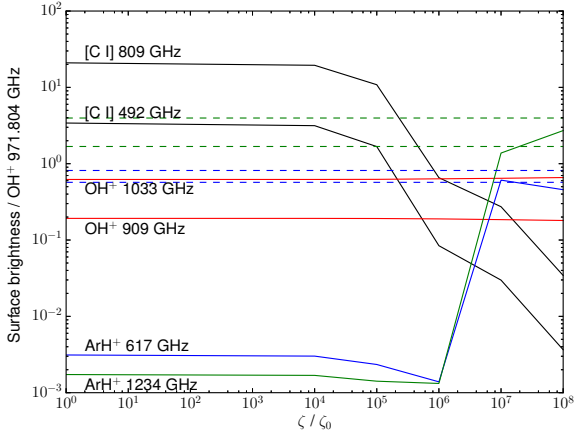}
  \caption{Line surface brightness ratios relative to OH$^+$ 971 GHz versus the cosmic ray ionisation rate $\zeta$ in units of $\zeta_0$, for models with $n_{\rm{H}} = 2 \times 10^4$ cm$^{-3}$ and a rate coefficient for ArH$^+$ + e$^-$ of $10^{-11}$ cm$^3$ s$^{-1}$, with the emissivity integrated up to $A_V = 0.01$, as for our D4Z7 and D4Z8 models (see Table~\ref{tab:models}). The dashed lines show upper and lower observational bounds for the ArH$^+$ transitions, from \citet{barlow2013}.}
  \label{fig:lineserate-11d2e4av1e-2}
\end{figure}

Figure~\ref{fig:lineserate-11d2e3} shows the relative line surface brightnesses versus $\zeta$ for a cloud density of $1900$ cm$^{-3}$, this time with the ArH$^+$ + e$^-$ reaction rate set to $10^{-11}$ cm$^3$ s$^{-1}$. For $\zeta \ge 10^6 \zeta_0$, the relative surface brightnesses of the two ArH$^+$ transitions have increased to values comparable with observations, although the absolute surface brightness of the OH$^+$ 971 GHz line is too high for these values of $\zeta$, as shown in Figure~\ref{fig:oh971av3}. Integrated only up to $A_V = 0.05$, the OH$^+$ surface brightness is within the observed range for all values of $\zeta$, while the relative strengths of the ArH$^+$ lines, shown in Figure~\ref{fig:lineserate-11d2e3av5e-2}, are broadly consistent with observation for $\zeta \ge 10^6 \zeta_0$, although for the $10^6 \zeta_0$ model the predicted [C I] 809 GHz surface brightness is too high. Figure~\ref{fig:lineserate-11d2e4av1e-2} shows the line surface brightness ratios integrated up to $A_V = 0.01$ for a density of $2 \times 10^4$ cm$^{-3}$, with the reduced dissociative recombination rate. Comparing with Figure~\ref{fig:linescrabd2e4av1e-2}, the ArH$^+$ line ratios are increased for $\zeta \ge 10^7 \zeta_0$. For these values of $\zeta$, all line surface brightnesses are consistent with observation, as both the relative line strengths and the OH$^+$ 971 GHz surface brightness fall within the observed range of values. The models with $n_{\rm{H}} = 2 \times 10^5$ and $2 \times 10^6$ cm$^{-3}$ do not reproduce the observed ArH$^+$ surface brightness ratios for any combination of $\zeta$ and $A_V$.

\section{Discussion}
\label{sec:discussion}

Our initial models, and models with an increased UV/X-ray radiation field, were unable to reproduce the observed Herschel SPIRE FTS line emission for any combination of parameters we tested. With a reduced rate for the ArH$^+$ dissociative recombination reaction, results broadly consistent with observation can be found for densities of $1900$ and $2 \times 10^4$ cm$^{-3}$ and high cosmic ray ionization rates, with the highest density models ruled out. Cosmic ray ionization rates of at least $\zeta = 10^7 \zeta_0$ are required to reduce the [C I] emission to below detectable levels. Selecting filament thicknesses which reproduce the observed OH$^+$ 971 GHz surface brightnesses, we restrict our analysis to four models which are entirely consistent with the Herschel SPIRE FTS observations, D3Z7, D3Z8, D4Z7 and D4Z8, with the model parameters given in Table~\ref{tab:models}.

\subsection{Physical properties}

\begin{figure}
  \centering
  \includegraphics[width=\columnwidth]{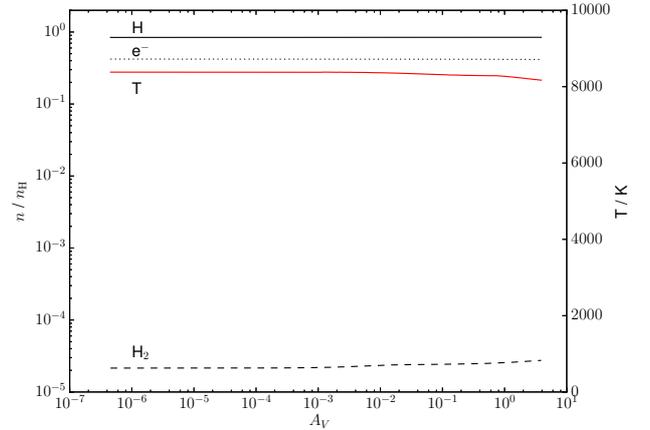}
  \caption{Temperature (red line) and abundance of H (solid black line), H$_2$ (dashed black line) and e$^-$ (dotted black line) relative to $n_{\rm{H}}$ versus $A_V$ for an $n_{\rm{H}} = 1900$ cm$^{-3}$, $\zeta = 10^7 \zeta_0$ model. These are the same parameters as model D3Z7, for which the cloud only extends up to $A_V = 0.05$.}
  \label{fig:properate-11d2e3z1e7}
\end{figure}

Figure~\ref{fig:properate-11d2e3z1e7} shows the gas temperature and abundances of H, H$_2$ and e$^-$ versus $A_V$ for an $n_{\rm{H}} = 1900$ cm$^{-3}$, $\zeta = 10^7 \zeta_0$ model, the values used for model D3Z7. Even for this case, with the lowest value of $\zeta$ of our preferred models, cosmic ray ionization dominates both the gas heating rate and the chemistry, so the filament properties do not change significantly with depth. The gas remains primarily atomic, with an ionized fraction of $\sim 0.4$ and a molecular fraction of $\sim 10^{-5}$, at a temperature of $\sim 8000$ K. For the $n_{\rm{H}} = 1900$ cm$^{-3}$, $\zeta = 10^8 \zeta_0$ model, there is no detectable change in any filament property with depth.

As the properties for our preferred models vary little with depth, we assume the values at the filament surface are representative for the whole filament and give the properties of our final models in Table~\ref{tab:prop}. We find that all models successfully reproducing the line surface brightness observations have H$_2$ abundances of order $10^{-5}$ and ionization fractions of at least $0.1$. We list the surface brightnesses of emission lines observed in the Crab Nebula for each model in Table~\ref{tab:emis}, along with observational upper and lower limits.

\begin{table*}
  \centering
  \caption{Gas temperatures and abundances of H, H$_2$, e$^-$, Ar$^+$ and ArH$^+$ for our final models.}
  \begin{tabular}{ccccccc}
    \hline
    Model & T / K & $n(\rm{H})$ / $n_{\rm{H}}$ & $n(\rm{H}_2)$ / $n_{\rm{H}}$ & $n(\rm{e}^-)$ / $n_{\rm{H}}$ & $n(\rm{Ar}^+)$ / $n_{\rm{H}}$ & $n(\rm{ArH}^+)$ / $n_{\rm{H}}$ \\
    \hline
    D3Z7 & $8381$ & $0.84$ & $2.1 \times 10^{-5}$ & $0.42$ & $6.5 \times 10^{-6}$ & $2.3 \times 10^{-8}$ \\
    D3Z8 & $11958$ & $0.53$ & $9.2 \times 10^{-6}$ & $1.24$ & $9.0 \times 10^{-6}$ & $8.1 \times 10^{-9}$ \\
    D4Z7 & $4745$ & $0.96$ & $5.0 \times 10^{-5}$ & $0.11$ & $2.2 \times 10^{-6}$ & $2.4 \times 10^{-8}$ \\
    D4Z8 & $8209$ & $0.84$ & $2.7 \times 10^{-5}$ & $0.42$ & $6.4 \times 10^{-6}$ & $2.9 \times 10^{-8}$ \\
    \hline
  \end{tabular}
  \label{tab:prop}
\end{table*}

\begin{table*}
  \centering
  \caption{Predicted surface brightnesses and observational limits for ArH$^+$, OH$^+$ and H$_2$ emission lines in the Crab Nebula, in erg cm$^{-2}$ s$^{-1}$ sr$^{-1}$. Observed values are taken from \citet{barlow2013} for ArH$^+$ and OH$^+$ and \citet{loh2011} for H$_2$.}
  \begin{tabular}{ccccc}
    \hline
    Model & OH$^+$ 971 GHz & ArH$^+$ J=1-0 & ArH$^+$ J=2-1 & H$_2$ $2.12$ $\mu$m \\
    \hline
    Observed & $3.4-10.3 \times 10^{-7}$ & $2.2-9.9 \times 10^{-7}$ & $1.0-3.8 \times 10^{-6}$ & $1.0-4.8 \times 10^{-5}$ \\
    D3Z7 & $6.0 \times 10^{-7}$ & $7.5 \times 10^{-7}$ & $9.9 \times 10^{-7}$ & $6.4 \times 10^{-7}$ \\
    D3Z8 & $6.2 \times 10^{-7}$ & $4.3 \times 10^{-7}$ & $1.0 \times 10^{-6}$ & $2.3 \times 10^{-7}$ \\
    D4Z7 & $4.1 \times 10^{-7}$ & $2.5 \times 10^{-7}$ & $5.6 \times 10^{-7}$ & $4.7 \times 10^{-7}$ \\
    D4Z8 & $7.8 \times 10^{-7}$ & $3.6 \times 10^{-7}$ & $2.1 \times 10^{-6}$ & $3.0 \times 10^{-7}$ \\
    AVD3Z7 & $1.0 \times 10^{-6}$ & $1.2 \times 10^{-6}$ & $1.7 \times 10^{-6}$ & $1.1 \times 10^{-6}$ \\
    AVD3Z8 & $1.0 \times 10^{-6}$ & $7.2 \times 10^{-7}$ & $1.7 \times 10^{-6}$ & $3.9 \times 10^{-7}$ \\
    AVD4Z7 & $3.4 \times 10^{-6}$ & $2.0 \times 10^{-6}$ & $4.7 \times 10^{-6}$ & $4.0 \times 10^{-6}$ \\
    AVD4Z8 & $6.5 \times 10^{-6}$ & $2.9 \times 10^{-6}$ & $1.8 \times 10^{-5}$ & $2.5 \times 10^{-6}$ \\
    \hline
  \end{tabular}
  \label{tab:emis}
\end{table*}

\subsection{H$_2$ vibration-rotation emission}

\begin{figure}
  \centering
  \includegraphics[width=\columnwidth]{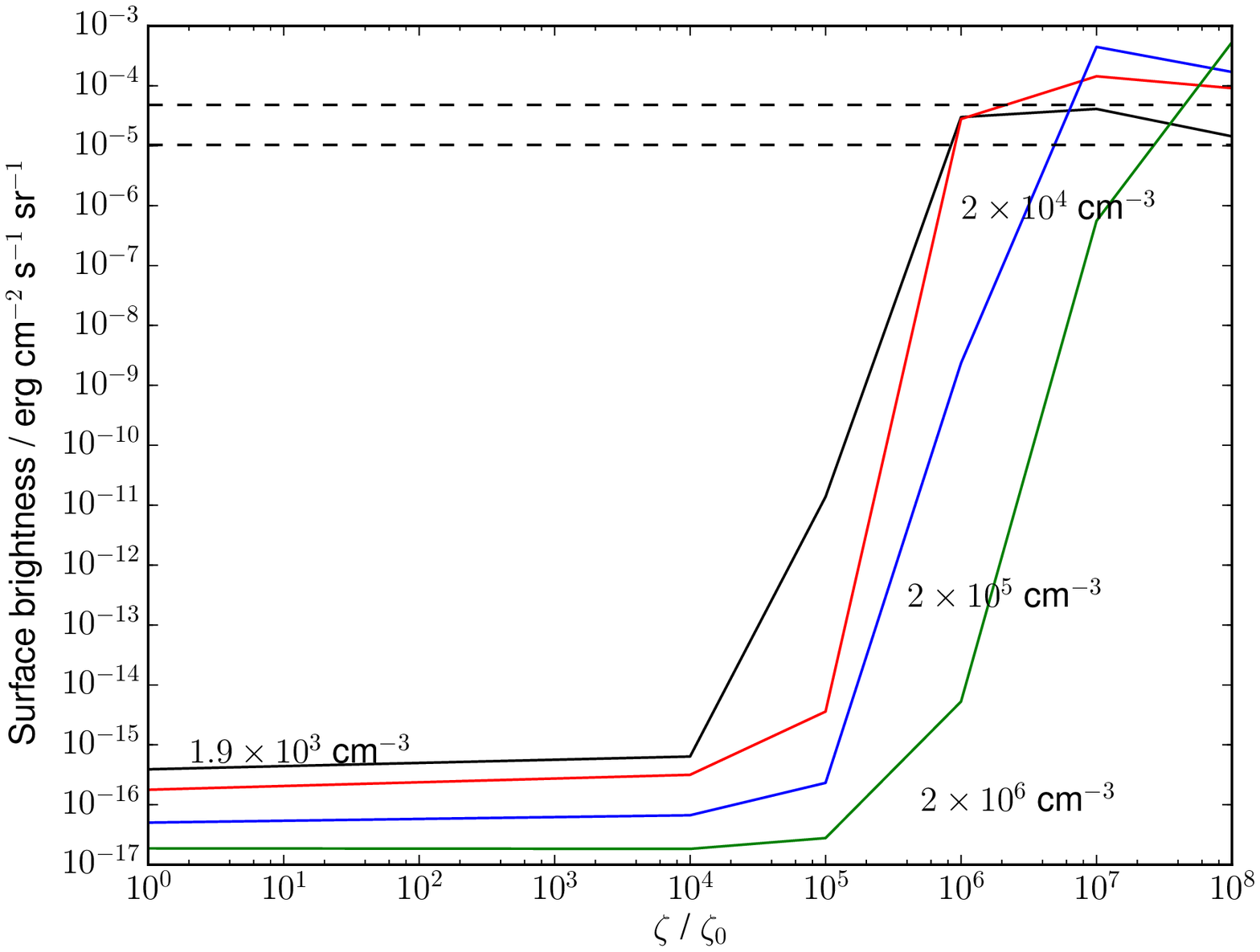}
  \caption{Absolute surface brightness integrated up to $A_V = 3$ of the H$_2$ $2.12$ $\mu$m transition versus cosmic ray ionisation rate $\zeta$ for models with $n_{\rm{H}} = 1900$ cm$^{-3}$ (black), $2 \times 10^4$ cm$^{-3}$ (red), $2 \times 10^5$ cm$^{-3}$ (blue) and $2 \times 10^6$ cm$^{-3}$ (green). The dashed lines show upper and lower observational bounds from \citet{loh2011}.}
  \label{fig:h2212av3}
\end{figure}

\begin{figure}
  \centering
  \includegraphics[width=\columnwidth]{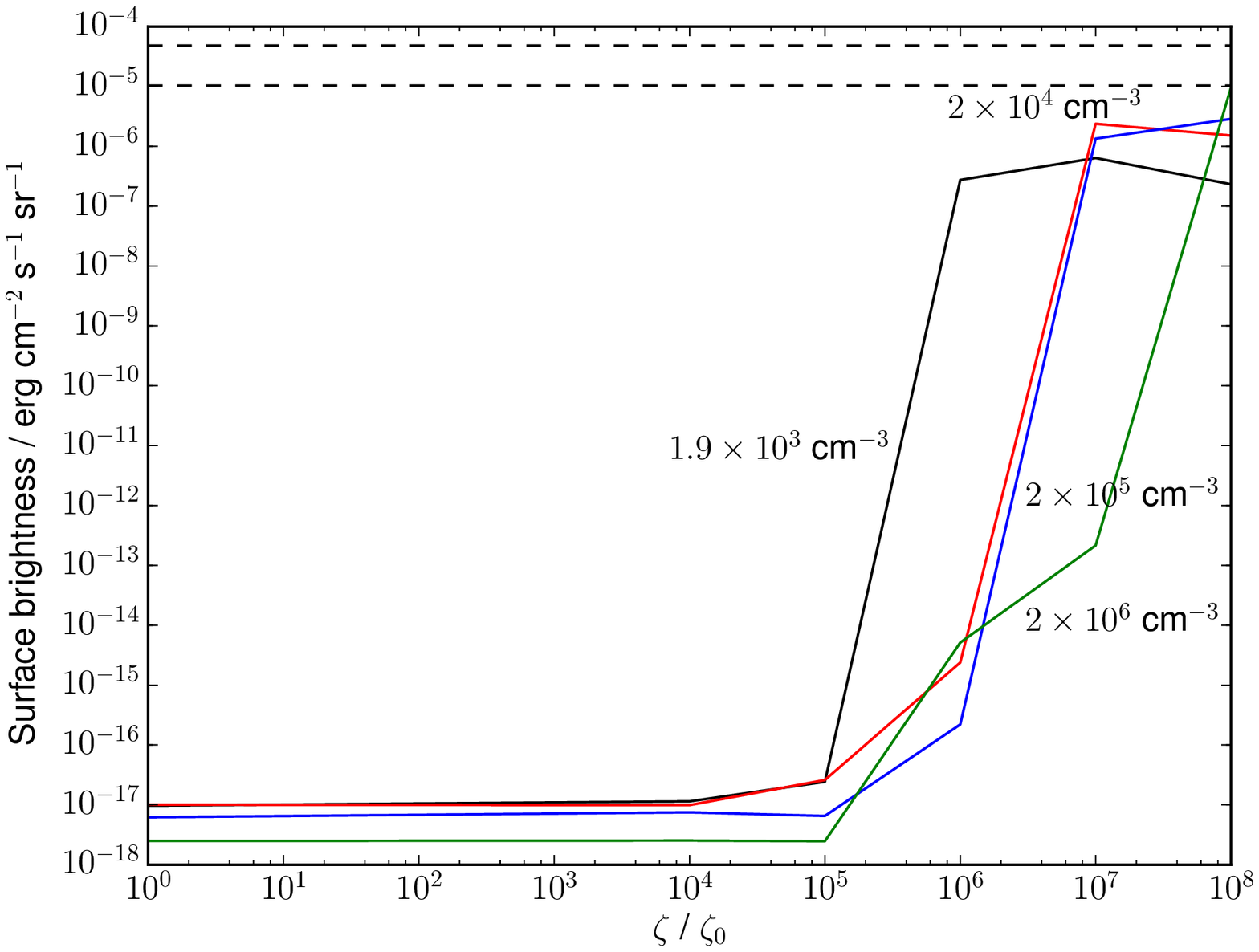}
  \caption{Absolute surface brightness integrated up to $A_V = 0.05$ (as for models D3Z7 and D3Z8) of the H$_2$ $2.12$ $\mu$m transition versus cosmic ray ionisation rate $\zeta$ for models with $n_{\rm{H}} = 1900$ cm$^{-3}$ (black), $2 \times 10^4$ cm$^{-3}$ (red), $2 \times 10^5$ cm$^{-3}$ (blue) and $2 \times 10^6$ cm$^{-3}$ (green). The dashed lines show upper and lower observational bounds from \citet{loh2011}.}
  \label{fig:h2212av5e-2}
\end{figure}

\citet{loh2010,loh2011} observed emission from the H$_2$ $2.12$ $\mu$m line from 55 knots in the Crab Nebula, with average surface brightnesses of $\sim 1-5 \times 10^{-5}$ erg cm$^{-2}$ s$^{-1}$ sr$^{-1}$. Figure~\ref{fig:h2212av3} shows the surface brightness of this line versus $\zeta$ for our models, integrated up to $A_V = 3$. Our $n_{\rm{H}} = 1900$ cm$^{-3}$ models give a 2.12 $\mu$m surface brightness consistent with observations, while for $n_{\rm{H}} = 2 \times 10^4$ cm$^{-3}$ the predicted surface brightnesses are somewhat higher than observed. However, these $A_V$ values give ArH$^+$ and OH$^+$ surface brightnesses far in excess of the observed values, as shown in Figure~\ref{fig:oh971av3}. For a filament thickness reduced to $A_V = 0.05$, shown in Figure~\ref{fig:h2212av5e-2}, the H$_2$ emission is reduced, with the models giving the highest values still an order of magnitude below observations. Our final models all predict H$_2$ surface brightnesses below the observed range, given in Table~\ref{tab:emis}.

\citet{loh2012} used H$_2$ vibration-rotation emission to derive the excitation temperature and density in Crab Nebula H$_2$ knots, finding a temperature range of $2000-3000$ K, and a lower limit of $n_{\rm{H}} = 2 \times 10^4$ cm$^{-3}$ for the density. While the limit on the density is consistent with two of our final models, D4Z7 and D4Z8, the gas temperatures for these models are higher than the derived excitation temperatures. Table~\ref{tab:h2lines} lists the surface brightnesses of the six H$_2$ vibrational transitions detected by \citet{loh2012}, relative to the 1-0 S(1) line, for a representative Crab Nebula knot from that paper and for our final PDR models. With the exception of D4Z8, all our models predict line ratios in agreement with the typical observed values. \citet{loh2012} also calculated the electron densities of the H$_2$ knots based on [S II] emission. Their typical electron abundances of $n(\rm{e}^-)/n_{\rm{H}} \approx 0.1$ are very close to the value of $0.11$ from our D4Z7 model, which also has the gas temperature closest to the measured H$_2$ exctiation temperature of all our final models.

\begin{table*}
  \centering
  \caption{H$_2$ vibrational line surface brightness relative to the 1-0 S(1) line, for Knot 51 from \citet{loh2012} and for our final models.}
  \begin{tabular}{ccccccc}
    \hline
    Line & 1-0 S(0) & 1-0 S(1) & 1-0 S(2) & 2-1 S(1) & 2-1 S(2) & 2-1 S(3) \\
    \hline
    Knot 51 & $0.23 \pm 0.04$ & $1.00 \pm 0.04$ & $0.52 \pm 0.09$ & $0.19 \pm 0.03$ & $<0.13$ & $<0.28$ \\
    D3Z7 & $0.16$ & $1.00$ & $0.44$ & $0.16$ & $0.06$ & $0.26$ \\
    D3Z8 & $0.16$ & $1.00$ & $0.44$ & $0.16$ & $0.06$ & $0.26$ \\
    D4Z7 & $0.18$ & $1.00$ & $0.43$ & $0.21$ & $0.09$ & $0.32$ \\
    D4Z8 & $0.18$ & $1.00$ & $0.48$ & $0.44$ & $0.20$ & $0.73$ \\
    \hline
  \end{tabular}
  \label{tab:h2lines}
\end{table*}

\citet{richardson2013} modelled a Crab Nebula H$_2$-emitting knot using \textsc{cloudy} \citep{ferland1998}, in order to explain the combination of near-infrared H$_2$ vibrational lines with optical emission from atomic and ionised gas. Their models required an additional heating source to reproduce the observed H$_2$ vibrational spectrum, which they assumed to be either energy input from shocks or turbulence, or from the ionising particle flux from the PWN. Their charged particle model corresponds to $\zeta = 6.1 \times 10^6 \zeta_0$, only slightly lower than the range of values we require. Their models assume the gas is in pressure balance, leading to a PDR region with the density increasing from $\sim 10^3$ to $10^5$ cm$^{-3}$. The gas temperature and H$_2$ and e$^-$ abundances are consistent with our final models where the densities overlap, although in the highest density regions the H$_2$ fractional abundance reaches $10^{-3}$, higher than any of our models. We note that in their models, these regions produce the majority of the H$_2$ $2.12$ $\mu$m emission, suggesting that an additional, denser gas component combined with our final models may solve the issue of underpredicting the H$_2$ surface brightness. However, as shown in Figure~\ref{fig:oh971av1e-2}, even a small amount of dense gas produces OH$^+$ 971 GHz emission in excess of the observed values, while Figure~\ref{fig:h2212av5e-2} shows that the H$_2$ $2.12$ $\mu$m surface brightnesses of these denser models are not significantly greater than for our $n_{\rm{H}} = 2 \times 10^4$ cm$^{-3}$ models.

\subsection{Far-infrared line emission}

\citet{gomez2012} listed Herschel PACS and ISO LWS fluxes for infrared fine structure emission lines from a number of species in the Crab Nebula, including [C II] and [O I], which are also modelled by \textsc{ucl\_pdr}. Table~\ref{tab:emis2} lists the predicted [O I] 63 $\mu$m/146 $\mu$m and [O I] 146 $\mu$m/[C II] 158 $\mu$m line ratios for our final models. The predicted ratios show reasonable agreement with the observed ratios from \citet{gomez2012}, although the D4Z7 and D4Z8 models predict [O I] 146 $\mu$m/[C II] 158 $\mu$m ratios twice as large as the highest value observed. However, we find that the predicted line surface brightnesses exceed the measured values by an order of magnitude. This overestimation may be due to the fact that our models do not treat any optical or UV line emission from higher ionization stages that would be produced by the high charged particle flux, and so require stronger emission from other species in order to balance the heating rate. [Fe II], which \citet{richardson2013} found to be the dominant coolant in the PDR region of their ionizing particles model, is also not included in our models.

\begin{table*}
  \centering
  \caption{Predicted and observed line surface brightness ratios for [O I], [C II] and HeH$^+$ emission lines in the Crab Nebula. Observed values are taken from \citet{gomez2012}.}
  \begin{tabular}{ccccc}
    \hline
    Model & $\frac{\rm{[O I]} 63 \mu\rm{m}}{\rm{[O I]} 146 \mu\rm{m}}$ & $\frac{\rm{[O I]} 146 \mu\rm{m}}{\rm{[C II]} 158 \mu\rm{m}}$ & $\frac{\rm{HeH}^+ 149 \mu\rm{m}}{\rm{[O I] 146} \mu\rm{m}}$ & $\frac{\rm{HeH}^+ 74.5 \mu\rm{m}}{\rm{[O I] 146} \mu\rm{m}}$ \\
    \hline
    Observed & $16.4-38.7$ & $0.125-0.323$ & - & - \\
    D3Z7 & $13.1$ & $0.303$ & $0.058$ & $0.055$ \\
    D3Z8 & $18.4$ & $0.213$ & $0.219$ & $0.221$ \\
    D4Z7 & $22.6$ & $0.650$ & $0.034$ & $0.034$ \\
    D4Z8 & $26.7$ & $0.582$ & $0.214$ & $0.250$ \\
    AVD3Z7 & $12.9$ & $0.305$ & $0.058$ & $0.055$ \\
    AVD3Z8 & $18.2$ & $0.214$ & $0.218$ & $0.220$ \\
    AVD4Z7 & $20.8$ & $0.660$ & $0.034$ & $0.034$ \\
    AVD4Z8 & $25.2$ & $0.585$ & $0.207$ & $0.247$ \\
    \hline
  \end{tabular}
  \label{tab:emis2}
\end{table*}

\subsection{Filament size and extinction}

The $A_V$ required to match the observed OH$^+$ 971 GHz surface brightness in our final models corresponds to filament thicknesses of $\sim 4 \times 10^{16}$ cm for $n_{\rm{H}} = 1900$ cm$^{-3}$ and $\sim 8 \times 10^{14}$ cm for $n_{\rm{H}} = 2 \times 10^4$ cm$^{-3}$. \citet{richardson2013} used a cloud thickness of $10^{16.5}$ cm, based on the observed width of the knot they modelled, close to the size of our $n_{\rm{H}} = 1900$ cm$^{-3}$ models. If the ionized region (which we do not attempt to model) is excluded, the PDR regions in their models are $\sim 5 \times 10^{15}$ cm in size, intermediate between our two values. \citet{loh2012} estimated the thickness of the H$_2$-emitting regions they observed to be of order $10^{14}$ cm, which is comparable to our $n_{\rm{H}} = 2 \times 10^4$ cm$^{-3}$ models.

\citet{grenman2017} used Hubble Space Telescope optical images of the Crab Nebula to determine the extinctions of several dusty globules observed as dark spots against the synchrotron background, obtaining values of $A_V = 0.20-0.34$. The globule sizes range from $6 \times 10^{15}$ cm to $3 \times 10^{16}$ cm, similar to the sizes of the H$_2$ emitting knots observed by \citet{loh2011}, although \citet{grenman2017} note that only $\sim 10 \%$ of their dusty globules are coincident with H$_2$ knots. While the observed globule sizes are similar to our $n_{\rm{H}} = 1900$ cm$^{-3}$ model cloud thicknesses, none of our preferred models match the extinctions found by \citet{grenman2017}. These models used a standard interstellar $A_V$/$N_{\rm{H}}$ ratio of $6.289 \times 10^{-22}$ cm$^2$ mag, whereas with the increased dust-to-gas ratio in the Crab Nebula the true value is likely to be higher.

We reran our final models with an $A_V$/$n_{\rm{H}}$ ratio increased by a factor of $3.6$ (the dust-to-gas mass ratio in the Crab relative to the standard ISM value) and an $A_V$ of $0.3$. The model parameters for these models, which we refer to as AVD3Z7, AVD3Z8, AVD4Z7 and AVD4Z8, are listed in Table~\ref{tab:models}. The temperature and abundances of the models were largely unchanged compared to models with the standard $A_V$/$n_{\rm{H}}$ and the same density and ionization rate, and the H$_2$ line ratios given in Table~\ref{tab:h2lines} were also essentially identical. We give the predicted line surface brightnesses in Tables~\ref{tab:emis} and \ref{tab:emis2}. The AVD4Z8 model predicts OH$^+$ and ArH$^+$ surface brightnesses significantly in excess of the observed values, while the other three models predict values either within or slightly above the observed range. The AVD4Z7 model additionally predicts an H$_2$ $2.12$ $\mu$m surface brightness only a factor of $2.5$ smaller than the observational lower limit, while also producing line ratios consistent with the observed values from \citet{loh2012}. The surface brightnesses of the [C II] and [O I] lines are higher than for the equivalent standard $A_V$/$n_{\rm{H}}$ models, while the line ratios are similar.

\subsection{Predicted HeH$^+$ rotational line emission}

HeH$^+$ has been predicted to form in ionized nebulae \citep{black1978,flower1979,roberge1982,cecchi1993}, as well as in the early Universe \citep{galli1998}, although the predicted intensities of both its rotational and vibrational transitions have been generally higher than the observational upper limts (e.g. \citet{moorhead1988,liu1997}). We incorporated HeH$^+$ into our chemical network using the reactions and rate coefficients from \citet{roberge1982}, with the exception of the He$^+$ + H reaction, for which we used the rate coefficient from \citet{zygelman1990}, and the HeH$^+$ + e$^-$ reaction, where we used the experimentally determined rate from \citet{yousif1989} (via \citet{cecchi1993}). We also neglected photodissociation of HeH$^+$, as our input SED from \textsc{mocassin} has negligible flux in the wavelength range beyond the Lyman limit relevant for the cross-section given by \citet{roberge1982}. We calculated the emissivity of the first three rotational transitions in the same way as for ArH$^+$, using electron impact excitation data from \citet{hamilton2016} and energy levels and A-coefficients from CDMS \citep{muller2001,muller2005}.

Our models predict HeH$^+$ J=1-0 $149.14$ $\mu$m and J=2-1 $74.57$ $\mu$m emission with similar surface brightnesses, from a few times $10^{-5}$ to a few times $10^{-4}$ erg cm$^{-2}$ s$^{-1}$ sr$^{-1}$, while the J=3-2 line is a factor of $\sim 20$ weaker. We give the line surface brightness ratios of the first two transitions compared to the [O I] 146 $\mu$m line in Table~\ref{tab:emis2} for our final models. For models with $\zeta = 10^7 \zeta_0$, the $149.14$ $\mu$m line is about 20 times weaker than the nearby [O I] 146 $\mu$m line, which is potentially detectable given the reported uncertainties in the [O I] line strengths, whereas for $10^8 \zeta_0$ the ratio increases to $\sim 0.2$ and the HeH$^+$ line should be clearly present. The $74.57$ $\mu$m ratios are similar to those for $149.14$ $\mu$m, or somewhat larger for the D4Z8 models, although the surface brightnesses are only around $1 \%$ of the value of the [O I] 63 $\mu$m line.

\subsection{Geometrical effects}

We treat the Crab Nebula filaments as homogenous one-dimensional slabs, with the PWN radiation field incident on one side and the observer on the other. In reality the Crab is a complex 3D structure, and all of these assumptions are likely to be incorrect to a greater or lesser extent. Each line of sight towards the Crab Nebula may intersect a number of gas components varying in density, temperature, radiation field strength and size, with the observed emission then being made up of multiple separate contributions. Additionally, the observed surface brightnesses are the average values over the area of the telescope beam, meaning that variations on smaller scales can be washed out. Our analysis, which assumes that the emission from all observed species originates from a single cloud, is obviously simplistic, but given the complexity of attempting to model multiple gas components using a 1D code, and our success at predicting molecular emission consistent with observation, we consider it acceptable for the purposes of this paper.

The effect of beam size on the observed surface brightnesses may resolve the issue of the excess [C II] and [O I] emission from our models. The dusty globules detected by \citet{grenman2017} have radii of $\leq 0.5$ arcsec, far smaller than the beam sizes in \citet{gomez2012} ($9.4\rm{''} \times 9.4\rm{''}$ for each PACS spaxel). The ratio of globule angular area to PACS spaxel area is $\sim 3 \times 10^{-3}$, which is low enough to explain the discrepancy between our models and the observations if each spaxel contains only a few globules, although unlike the moleuclar emission, there is no reason to assume that the far-infrared lines only originate in dense, shielded clumps. The Hershel SPIRE FTS beam size is also much larger than the typical angular sizes of the H$_2$ knots from \citet{loh2011}, so assuming that the ArH$^+$ and OH$^+$ emission also originates in these knots, a higher model surface brightness would be necessary to reproduce observed values. If this was obtained by increasing the filament $A_V$ value, it would also increase the H$_2$ $2.12$ $\mu$m surface brightness, potentially removing the issue of our models predicting lower values than those from \citet{loh2011}.

\subsection{Time-dependent effects}

Our PDR models assume that the gas is in equilibrium, with the chemistry and thermal balance both in steady state. However, as the Crab Nebula is a relatively young object ($\sim 950$ yr), this assumption is likely to be invalid. \citet{richardson2013} noted this issue and found that the timescale for the formation of H$^-$, which controls the formation of H$_2$, exceeds the age of the Crab in their models, indicating that time-dependent calculations are required. However, modelling the chemical evolution of the Crab Nebula from supernova to its current state is a complex problem well beyond the scope of this paper. We therefore focus on the timescales for reactions relevant to the formation of the molecules.

Our models find a minimum fractional ionization of Ar$^+$/Ar$^0$ $\approx 0.2$, while for lower densities and higher values of $\zeta$ the value increases. The main formation route for ArH$^+$, Ar$^+$ + H$_2$, has a rate of $\sim 10^{-9}$ cm$^3$ s$^{-1}$ \citep{rebrion1989}, so for an H$_2$ density of $\sim 1$ cm$^{-3}$ (for $n_{\rm{H}} = 2 \times 10^4$ cm$^{-3}$) the formation timescale is $\sim 10^9$ s, or around 30 years, much less than the age of the Crab. This suggests ArH$^+$ can form efficiently from Ar$^+$ under Crab Nebula conditions. However, this assumes an equilibrium abundance of H$_2$, which also has to form from the presumably initially fully ionized supernova ejecta. Using the values from our models, we find a similar timescale to \citet{richardson2013} of $\sim 10^4$ yr for the formation of H$_2$ by associative detachment, which is significantly longer than the age of the Crab. Although this suggests our value for the formation timescale of ArH$^+$ is an underestimate, we note that even with an order of magnitude increase our timescale is still shorter than the age of the Crab, and that earlier in the history of the Crab Nebula the gas was presumably both denser and more highly ionized, which would tend to shorten both formation timescales. Our models predict less H$_2$ emission than is actually observed in the Crab, raising the prospect that we may in fact be underestimating, rather than overestimating, the H$_2$ abundance. It therefore seems plausible that enough ArH$^+$ could have formed up to the present age of the Crab Nebula to account for the observed line emission.

In our chemical network, HeH$^+$ is formed by the reaction of He with H$^+$ and H$_2^+$, and the reaction of He$^+$ with H. The He$^+$/He ratio ranges from $0.03$ to $0.75$ depending on the model. Calculating the typical timescales of all these reactions (the time for He/He$^+$ to be converted to HeH$^+$), we find the dominant reaction is He$^+$ + H for all our preferred models, with a timescale shorter than the others by several orders of magnitude. However, even for this reaction the lowest formation timescale is at least a few $\times 10^3$ yr, and this value occurs for the model with the lowest He$^+$ abundance. It therefore appears that HeH$^+$ should form at a much slower rate and on a longer timescale than ArH$^+$.

\section{Conclusions}
\label{sec:conclusion}

We have modelled a Crab Nebula filament using a combination of photoionisation and PDR codes, varying the density and cosmic ray ionisation rate to create a grid of models. We find that a high energy particle ionisation rate of $>10^{7} \zeta_0$ is required in order to explain the lack of detected [C I] and CO emission in the Herschel SPIRE FTS spectra reported by \citet{barlow2013}, while the observed line surface brightness ratios of ArH$^+$ to OH$^+$ can only be reproduced with these high ionisation rates and a reduced electron recombination rate for ArH$^+$. This is consistent with both experimentally determined reaction rates and theoretical estimates of the charged particle flux in the Crab. The Herschel observations can be entirely explained by models with densities between $n_{\rm{H}} = 1900$ and $2 \times 10^4$ cm$^{-3}$, while models with higher densities fail to reproduce the observed ArH$^+$ to OH$^+$ line ratios. Our models agree with the conclusion of \citet{richardson2013} that the gas in the Crab Nebula filaments is primarily atomic, even in the regions with strong H$_2$ emission.

Using an increased $A_V$/$N_{\rm{H}}$ ratio more appropriate for the Crab Nebula, and a total extinction within the range found by \citet{grenman2017} for Crab Nebula dusty globules, we find that models with $n_{\rm{H}} = 1900$ cm$^{-3}$ reproduce the observed ArH$^+$ and OH$^+$ surface brightnesses measured by \citet{barlow2013}, while underpredicting the H$_2$ surface brightness compared to \citet{loh2010,loh2011,loh2012}. Our $n_{\rm{H}} = 2 \times 10^4$ cm$^{-3}$, $\zeta = 10^7 \zeta_0$ model produces surface brightnesses within a factor of a few of observations for ArH$^+$, OH$^+$ and H$_2$, while the $n_{\rm{H}} = 2 \times 10^4$ cm$^{-3}$, $\zeta = 10^8 \zeta_0$ model predicts significantly stronger ArH$^+$ and OH$^+$ emission, as well as failing to reproduce the H$_2$ vibration-rotation line ratios. All models predict [C II] and [O I] line emission in excess of the values listed by \citet{gomez2012}, although surface brightness ratios are comparable with observed values. We suggest that these dusty globules are the source of most of the observed molecular emission in the Crab Nebula, with denser gas accounting for the $\sim 10 \%$ of globules associated with H$_2$ vibrational emission. Although our models predict HeH$^+$ rotational emission above detection thresholds, the formation timescales for this molecule are much longer than the age of the Crab Nebula, lowering the expected HeH$^+$ line strengths. The formation timescale for ArH$^+$, in contrast, is significantly shorter, suggesting that it can be formed efficiently under the conditions prevalent in the Crab Nebula.

\section*{Acknowledgements}

FP is supported by the Perren fund and IMPACT fund. The authors would like to acknowledge the work of Dr. Tom Bell, who was responsible for rewriting and substantially upgrading \textsc{ucl\_pdr}.




\bibliographystyle{mnras}
\bibliography{argon} 


\appendix

\section{Cosmic ray-induced dissociation reactions}
\label{sec:crreacs}

The chemical network which we used does not include cosmic ray-induced dissociation of OH$^+$ or ArH$^+$ (either directly or by secondary photons). If we assume that the rates for these reactions are of the same order of magnitude as for H$_2$ destruction ($\sim 10^{-17}$ s$^{-1}$ for $\zeta = \zeta_0$), the reaction rate per molecule would be of order $10^{-9}$ s$^{-1}$ for the $10^{8} \zeta_0$ models. At the typical electron densities of $n_{\rm{e}} \approx 10^3$ cm$^{-3}$ for these models, then even for the lower electron dissociative recombination rate that we used ($10^{-11}$ cm$^3$ s$^{-1}$) the electron recombination destruction rate per molecule for this species will be $10^{-8}$ s$^{-1}$, an order of magnitude higher. This ignores the effect of secondary photons, but both molecules are relatively resistant to photodissociation by UV radiation - the primary photodissociation rate for ArH$^+$ is $10^2$ times smaller than that for OH, and if the photodissociation rate from secondary photons scales similarly it will only be a factor of a few larger than the primary rate. Therefore even for this extreme example, cosmic ray induced destructions are unlikely to be dominant for ArH$^+$ and OH$^+$.

\bsp	
\label{lastpage}
\end{document}